\documentclass{aa}

\usepackage{newtxtext}
\usepackage[T1]{fontenc}

\DeclareRobustCommand{\VAN}[3]{#2}
\let\VANthebibliography\thebibliography
\def\thebibliography{\DeclareRobustCommand{\VAN}[3]{##3}\VANthebibliography}

\usepackage{graphicx}
\usepackage{natbib}
\usepackage{amssymb}
\usepackage{amsmath}
\usepackage{xspace}
\usepackage{dirtytalk}
\usepackage{placeins}
\usepackage[dvipsnames]{xcolor}
\usepackage{comment}
\usepackage{multirow}
\usepackage[colorlinks=true,     linkcolor=blue, citecolor=blue, filecolor=blue, urlcolor=blue]{hyperref}

\usepackage[normalem]{ulem}
\usepackage{soul}
\usepackage[varg]{txfonts}

\newcommand{\rb}[1]{{\color{black}  #1}} 
\newcommand{\is}[1]{{\color{black}  #1}} 

\newcommand{\isrev}[1]{{\color{black}  #1}} 
\newcommand{\isBlasi}[1]{{\color{black}  #1}} 

\defcitealias{2022MNRAS.516..492B}{Paper I}

\graphicspath{{./}{figures/}}

\begin{document}
%
%
\providecommand{\nat}{Nature}  
\providecommand{\aanda}{A\&A}  
\providecommand{\aaps}{A\&AS}  
\providecommand{\aap}{A\&A}  
\providecommand{\aapr}{A\&A Rev.}  
\providecommand{\aj}{AJ}      
\providecommand{\apj}{ApJ}      
\providecommand{\apjl}{ApJL}   
\providecommand{\apjs}{ApJS}   
\providecommand{\mnras}{MNRAS} 
\providecommand{\memras}{Mem.~RAS} 
\providecommand{\newa}{NewA}   
\providecommand{\na}{NewA}
\providecommand{\jcp}{JCP}     
\providecommand{\rmxaa}{RMxAA} 
\providecommand{\pasj}{PASJ}   
\providecommand{\pasp}{PASP}   
\providecommand{\apss}{AP\&SS} 
\providecommand{\araa}{ARA\&A} 
\providecommand{\bain}{Bull.~Astron.~Inst.~Netherlands} 
\providecommand{\physrep}{Physics Reports}
\providecommand{\ssr}{Space Science Reviews}
\providecommand{\pre}{Physical Review E}
\providecommand{\jgr}{Journal of Geophysics Research}
\providecommand{\prl}{Phys.~Rev.~Lett.}
\providecommand{\prd}{Phys.~Rev.~D}%

\title{How to turn a Supernova into a PeVatron}

\author{R. Brose$^{1,2,3}$\thanks{E-Mail: robert.brose@desy.de} 
    I. Sushch$^{4,5,6,7,8}$
    J. Mackey$^{3}$}
\institute{
Institute of Physics and Astronomy, University of Potsdam, 14476 Potsdam-Golm, Germany
\and
School of Physical Sciences and Centre for Astrophysics \& Relativity, Dublin City University, Glasnevin, D09 W6Y4, Ireland.
\and
Dublin Institute for Advanced Studies, Astronomy \& Astrophysics Section,  DIAS Dunsink Observatory, Dublin D15 XR2R, Ireland
\and
Centro de Investigaciones Energ\'eticas, Medioambientales y Tecnol\'ogicas (CIEMAT), E-28040 Madrid, Spain
\and
Gran Sasso Science Institute, Via F.Crispi 7, 67100 L’Aquila, Italy
\and
INFN-Laboratori Nazionali del Gran Sasso, Via G. Acitelli 22, Assergi (AQ), Italy
\and
Centre for Space Research, North-West University, 2520 Potchefstroom, South Africa
\and
Astronomical Observatory of Ivan Franko National University of Lviv, Kyryla i Methodia 8, 79005 Lviv, Ukraine}

\date{Received ; accepted}
\authorrunning{R. Brose et al.}

\abstract
{It is important to determine which Galactic cosmic-ray sources can accelerate particles to the knee of the cosmic ray spectrum at a few PeV, and in particular whether supernova remnants may contribute.
Current models for particle acceleration in very young remnants assume the circumstellar material consists of smooth, freely expanding winds.
There is strong evidence that some supernovae expand into much denser circumstellar material including dense shells ejected by eruptions shortly before explosion.}
{We investigate the effects of dense circumstellar shells on particle acceleration in supernova shocks during the first few years post-explosion, to quantify whether such interaction supernovae may act as PeVatrons.}
{We used the \textsc{pion} code to model the circumstellar medium around Luminous Blue Variables after having a brief episode with a mass-loss rate of up to $\dot{M}=2M_\odot/$yr. Consequently, we performed spherically symmetric 1-D simulations using our time-dependent acceleration-code \textsc{RATPaC} in which we simultaneously solve the transport equations for cosmic-rays, magnetic turbulence, and the hydrodynamical flow of the thermal plasma in the test-particle limit.}
{We find that the interaction with the circumstellar shells can significantly boost the maximum energy by enhancing particle escape during the onset of the shock-shell interaction followed by the reacceleration of the shock  propagating into a medium with a pre-amplified field. Early interactions boost the maximum energy to a greater degree and interactions within the first 5 months after explosion can increase $E_\text{max}$ to more then $1\,$PeV.} 
{}

\maketitle


\section{Introduction}

For almost hundred years supernova remnants (SNRs) have been extensively studied and discussed as prime sources of Galactic Cosmic Rays (CRs) with energies up to the knee feature in the CR spectrum at a few PeV \citep[see e.g.][]{1934PNAS...20..259B, 1987PhR...154....1B}. However, despite three decades of remarkable development in gamma-ray astronomy there is still no undisputed observational evidence that SNRs can accelerate particles above $\sim100$~TeV.
Young SNRs such as Tycho and Casiopeia A, that were expected to be effective particle accelerators, show even lower cutoff energies \citep{2020ApJ...894...51A}. 

Particle acceleration and high-energy radiation at shocks depend on the shock speed, the mass flux through the shock (i.e. pre-shock density), and the ambient magnetic field, so we may expect that a structured CSM will significantly modify the predicted CR population and non-thermal lightcurve of a supernova.
\citet{MurThoLac11} and \citet{MurThoOfe14} were the first to investigate this, finding that SNe with strong circumstellar interaction could accelerate particles beyond a PeV and be detectable in gamma-rays out to 200 Mpc, with bright accompanying non-thermal radio emission.

It is possible that only very young SNRs evolving in dense environments may satisfy the necessary conditions to accelerate particles to PeV energies. Driving of magnetic turbulence to the relevant scales through excitation of the non-resonant streaming instability \citep{2004MNRAS.353..550B} requires high CR currents that are possible only during  the initial $\sim 20$~years of the SNR evolution \citep{2013MNRAS.431..415B}, while high initial shock velocity and high ambient density further boost particle acceleration \citep{2018MNRAS.479.4470M, 2020MNRAS.494.2760C}. 
Recent active research in this field resulted in a common conclusion that only progenitor stars with an extremely dense circumstellar medium (CSM), produced by periods of high mass-loss rates with slow wind speed, have the required shock conditions to reach PeV energies \citep{2018MNRAS.479.4470M, 2020APh...12302492C, 2020MNRAS.494.2760C, 2021ApJ...922....7I, 2023ApJ...958....3D}.
If true, this means that only a relatively rare subset of supernova explosions can contribute anything to the flux of Galactic CRs around the knee at 3\,PeV.

In our recent paper, \citet[][hereafter \citetalias{2022MNRAS.516..492B}]{2022MNRAS.516..492B}, we investigated particle acceleration and non-thermal emission from SNRs associated with explosions of the Luminous Blue Variable (LBV) and Red Supergiant (RSG) progenitors during the first 20 years of their evolution, for explosions into smooth $1/r^2$ density profiles appropriate for stellar winds without variability.
We assessed the problem by numerical simulations with the \textbf{R}adiation \textbf{A}cceleration \textbf{T}ransport \textbf{Pa}rallel \textbf{C}ode (\textsc{RATPaC}) software extensively described in previous works \citep{Telezhinsky.2012a, Telezhinsky.2013, 2016A&A...593A..20B, 2019A&A...627A.166B, 2018A&A...618A.155S}.
\textsc{RATPaC} uses the kinetic approach to solving the transport equation for particles in time and space simultaneously with the solution of the hydrodynamic equations of the thermal plasma. It also solves the transport equation for the energy density in magnetic turbulence that is self-generated by accelerated particles. This approach allows us to self-consistently calculate the spectrum of accelerated particles and estimate the maximum energy achievable in the acceleration process, naturally accounting for the time required to build up the turbulence.
We showed that for the most favorable values of the stellar wind parameters, the maximum energy saturates at about $600$~TeV for the LBV case. More common densities of the circumstellar medium (CSM) facilitate acceleration only up to $200\,$TeV and $70\,$TeV for a LBV and RSG progenitor, respectively.
These results agree with other works that predict acceleration up to at most a few PeV for the most optimistic scenarios \citep[see e.g.][]{2018MNRAS.479.4470M}. Differences in the estimated  maximum energy could be attributed to assumptions implemented for the microphysics of the turbulence generation that vary from one study to another.
The feasibility of these assumptions will be discussed further in this paper.  

So far all the studies, including \citetalias{2022MNRAS.516..492B}, considered the expansion of the SNR into a smooth featureless wind while reality could strongly differ from this assumption.
While many aspects of mass loss are poorly understood during late evolutionary stages of massive stars, stars may undergo rapid changes in radius with associated change in mass-loss rate and wind speed \citep[e.g.][]{Lan12}.
The extreme example are the LBVs that undergo cyclic variations in radius of a factor of up to 10 on a timescale of a few years \citep{1994PASP..106.1025H, 2021A&A...647A..99G}, resulting in repeated episodes of mass loss through fast and then slow winds.
Some LBVs also undergo giant eruptions where a few solar masses of material are ejected on a timescale of a year, expanding with $\sim100$\,km\,s$^{-1}$, the most famous example being $\eta$ Carinae \citep{Smi14}.
At least some of these giant eruptions occur shortly before core-collapse, as in the case of SN 2009ip \citep{2013MNRAS.430.1801M}.
Binary interaction has emerged as a key aspect of the evolution of massive stars \citep[e.g.][]{2014ApJ...782....7D} with strong consequences for the CSM due to rapid and asymmetric mass loss through a common-envelope or Roche-lobe-overflow phase.
Common envelope followed by a stellar merger is a possible explanation for the dramatic 3-ring structure revealed in the CSM of SN1987A \citep{MorPod06, 2019MNRAS.482..438M}.
Stellar mergers appear to generate strong magnetic fields in massive stars \citep{2019Natur.574..211S}, which will in turn produce a strongly magnetised CSM.
Through these processes, stellar evolution from single and binary stars may produce a highly structured CSM around a star at the time of explosion, with sufficiently high density and inertia to affect the hydrodynamic expansion of the supernova shock in the first months and years of expansion.



In this paper, using the same approach as in \citetalias{2022MNRAS.516..492B}, we explore such SN-CSM interactions with spherically symmetric calculations and their impact on the maximum energy achievable in the acceleration process. The resulting emission both thermal and non-thermal will be examined in a follow-up publication.
Section~\ref{sec:methods} describes the model setup and introduces our assumptions for density and magnetic field of pre-supernova circumstellar matter. We also further discuss implications of certain aspects of our numeric simulations and assumptions that they rely upon. In section~\ref{sec:results} we describe the time-evolution of the accelerated particle spectra. Section~\ref{sec:conclusions} presents our conclusions.

\section{Model setup}
\label{sec:methods}

The numerical setup of simulations as well as basic assumptions and equations that are solved are the same as in \citetalias{2022MNRAS.516..492B}. We use the \textsc{RATPaC} code to combine a kinetic treatment of the CRs with a thermal leakage injection model, a fully time-dependent treatment of the magnetic turbulence, and a \textsc{PLUTO}-based hydrodynamic calculation. The essential ansatz to describe the evolution of the CR-distribution is solving the kinetic diffusion-advection equation for CR-transport \citep{Skilling.1975a}
\begin{align}
    \frac{\partial N}{\partial t} =& \nabla(D_r\nabla N-\mathbf{u} N)\nonumber\\
 &-\frac{\partial}{\partial p}\left( (N\dot{p})-\frac{\nabla \cdot \mathbf{ u}}{3}Np\right)+Q
\label{CRTE}\text{ , }
\end{align}
where $D_r$ denotes the spatial diffusion coefficient, $\textbf{u}$ the advective velocity, $\dot{p}$ energy losses and $Q$ the source of thermal particles.

For detailed description of the setup we refer the reader to \citetalias{2022MNRAS.516..492B}, while here we focus on the only difference in the setup compared to our previous work which is the structure of the CSM into which the SNR is expanding. Additionally we further discuss and analyse the assumptions we adopt for the microphysics of the self-generated turbulence and their impact on the results.

\subsection{Circumstellar medium}
In this work, we focus on the most extreme cases of possible CSM shells.
We identified LBV eruptions, where the mass-loss rate can reach values as high as $1\, \mathrm{M}_\odot\,$yr$^{-1}$ for periods of a few years as the case where we expect the densest CSM \citep{Smi14}.
However, Type IIn SNe that are observationally associated with LBVs make up only $\sim 5\,$\% of the CCSN rate \citep{2023A&A...670A..48C}.


\label{sec:LBV-wind}

The simplest model for an LBV eruption is to impose a dramatically increased mass-loss rate for a short period of time.
We modelled an evolving stellar wind with the astrophysical fluid-dynamics code \textsc{pion} \citep{2021MNRAS.504..983M} in spherical symmetry \citep{MacMohGva14}, by simulating the expansion of a stellar wind in three phases. 
The stellar wind is injected at its terminal velocity, $v_\infty$, with density set by the mass-loss rate $\dot{M}$ and the continuity equation.
Wind pressure is set by assuming adiabatic expansion from the stellar surface, but its exact value is unimportant for large Mach number winds.
The wind injection region is typically the first two grid cells closest to the origin.
The first phase has mass-loss rate $\dot{M} = 10^{-4}\,\mathrm{M}_\odot\,\mathrm{yr}^{-1}$ and wind terminal velocity $v_\infty=100$\,km\,s$^{-1}$ and lasts for 1800 years; the second (outburst) phase has $\dot{M} = 1\,\mathrm{M}_\odot\,\mathrm{yr}^{-1}$ and $v_\infty=100$\,km\,s$^{-1}$ and lasts for 2 years; the third is the same as the first but we integrate for 10\,000 years so that the shell expands to about 1\,pc radius.
The density structure of the shell is then used to construct the circumstellar shell in the \textsc{RATPaC} simulations as described below. 

In \citetalias{2022MNRAS.516..492B} we modeled the CSM density profile following a $\propto1/r^2$ distribution applicable for the free wind. Here we additionally introduce shells of enhanced mass following a Gaussian distribution,
\begin{align}
 \rho(r) = \frac{M_\text{shell}}{4\pi r^2\sqrt{2\pi d_\text{shell}^2}}\exp\left(-\frac{(r-R_\text{shell})^2}{2 d_\text{shell}^2}\right)
 \text{,}\label{eq:ShellProfile}
\end{align}
where $M_\text{shell}$ is the shell mass, $d_\text{shell}$ the shell thickness and $R_\text{shell}$ the radial position of the shell. The parameters for the free winds and the corresponding shells are given in Table~\ref{tab:ProgenitorModels}.

\begin{table*}[h]
  \centering
  \caption{Parameters for the progenitor stars winds and initial remnant sizes \citep{2014MNRAS.440.1917D, 2021A&A...647A..99G, 2021ApJ...908...75B}.
  Columns 2-3 give the mass-loss rate (2) and wind velocity (3) (assumed constant for the free wind), and colums 4-6 give the initial radius of the SN ejecta at the start of the simulation (4), the total ejecta mass of the SN (5) and the parameter $n$ determining the radial dependence of the ejecta (6). Colums 7-9 give the shell mass (7), shell thickness (8) and position of the shell (9) (see text for details).
  }
  \label{tab:ProgenitorModels}
  \begin{tabular}{c|c c | c c c | c c c}
     & \multicolumn{2}{c|}{Smooth Wind} & \multicolumn{3}{c|}{SN ejecta} & \multicolumn{3}{c}{Circumstellar Shell} \\
    Model & $\dot{M} [M_\odot/$yr] & $V_\mathrm{w}$ [km/s]  & $R_\text{ej}$ [cm] & $M_\text{ej}$ [$M_\odot$] & $n$ & $M_\text{shell}$ [$M_\odot$] & $d_\text{shell}$ [mpc] & $R_\text{shell}$ [mpc]\\
    \hline
    LBV 0.1 & \multirow{6}{*}{$10^{-2}$} & \multirow{6}{*}{100} & \multirow{6}{*}{$1.2\times10^{14}$} & \multirow{6}{*}{10} & \multirow{6}{*}{10} & \multirow{6}{*}{2} & \multirow{6}{*}{0.2} & 1.0\\
    LBV 0.3 &  &  &  &  &  &  &  & 3.0\\
    LBV 0.5 &  &  &  &  &  &  &  & 4.3\\
    LBV 0.75 &  &  &  &  &  &  &  & 6.2\\
    LBV 1.0 &  &  &  &  &  &  &  & 8.0\\
    LBV 10 &  &  &  &  &  &  &  & 61.0\\
  \end{tabular}
\end{table*}

\begin{figure}
    \centering
    \includegraphics[width=0.49\textwidth]{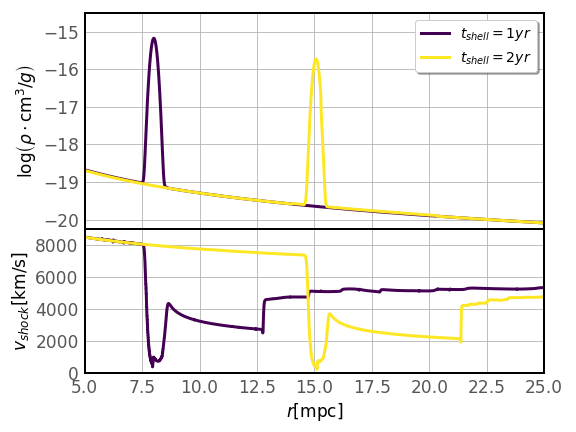}
    \caption{\rb{{\bf Upper panel: }Density profile of the CSM for the two of our LBV progenitor models. The shell parameters were obained from  \textsc{pion} simulations. The parameters of the shells used in \textsc{RATPaC} are  listed in table \ref{tab:ProgenitorModels}. {\bf Lower panel: } Shock velocity of the models as a function of the shock-position. Upward steps in the shock velocity are from interactions of the forward shock with reflected shocks catching-up from behind.
    }}
    \label{fig:Shells}
\end{figure}

We obtained the shell parameters by fitting the results of the stellar-wind simulations described before with the Gaussian profiles described  by equation~(\ref{eq:ShellProfile}). An example for the CSM structure as \rb{used in the \textsc{RATPaC}} simulations is shown in Figure \ref{fig:Shells}.

The total mass of material in the CSM is important not only for the shock dynamics but also for absorption in the ambient medium, for instance of X-rays and radio waves. 
It has to be noted that we ignore the effects of radiative cooling in our calculations. Strictly speaking, especially around the reverse-shock, the shock will be radiative initially but the intense photon-fields would require the use of a different equation of state for the plasma. The interactions with the dense shells will push the shock-dynamics into the radiative regime for a brief period even for the forward-shock for our LBV cases with the earliest interactions. However, resolving the changed dynamics of radiative shocks would be very challenging, and the particle injection algorithms would need significant modification if the post-shock cooling length could not be spatially resolved.


\rb{We model the large-scale magnetic field in the wind identical to \citetalias{2022MNRAS.516..492B}. Hence, the field is described as 
\begin{align}
    B(r)&= B_*\frac{R_*}{r}, 
\end{align}
where $B_*$ is the surface magnetic field and $R_*$ the radius of the progenitor. We assumed a constant ratio of $B_*(R_*/R_\odot=1000\,$G for both progenitor cases. In this work, we did not account for any potential compression of the magnetic field in the shells - studying this would require a more profound understanding of the variation of the surface-field of LBVs and RSGs throughout their lifetime. Additional structure in the magnetic field - especially compression in the RSG case - potentially introduces effects such as those described in \cite{2022ApJ...926..140S}, whereas this work aims to understand the impact of the density structure of the CSM.} 

\subsection{Self-generation of magnetic turbulence}\label{sec:Turb}
Any treatment of diffusive shock acceleration (DSA) needs a prescription for the scattering of CRs, as they need to be returned to the shock from the downstream region in order to undergo repeated shock-crossings. The various models on the market take different approaches in describing the scattering or diffusion of CRs and we will briefly review the current status of the field, to place our ansatz in the appropriate context.

The only numerical method that is able to capture all the microphysics involved is particle-in-cell simulation. However, to this point the computational costs are too large to simulate time-scales beyond a few ion-cyclotron times, $\omega_i$, and usually need to use reduced ion-to-electron mass-ratios \citep[e.g.][]{2009ApJ...690..244A, 2019ApJ...878....5B, 2019ApJ...885...10B, 2022RvMPP...6...29A}. Hybrid simulations, that treat electrons as a fluid are able to cover longer time-periods, however the simulated time-scales are still of the order of hours and it has at least be questioned that the observed acceleration is yet fully DSA-governed \citep{2014ApJ...783...91C, 2014ApJ...794...46C, 2014ApJ...794...47C, 2020ApJ...905....2C}.

All other approaches to tackling the shock acceleration problem rely on simplifications of the underlying microphysics in one or multiple aspects.

\subsubsection{Scattering problem}
One central aspect is what we will call the ``scattering problem'' in the following. CRs have to be returned to the shock in order to participate in the acceleration process. Fermi's initial description of DSA just relied on the assumption that magnetic structures or mirrors exist, that scatter the particles isotropically in the rest-frame of the local plasma \citep{1949PhRv...75.1169F}. By now, it is established, that CRs are scattered by (magnetic) turbulence in the plasma, although which wave-modes exactly are responsible for the scattering is still subject to ongoing research. Among the types of turbulence that are discussed are Alfven-waves \citep[e.g.][]{Skilling.1975a}, Bell's mode waves \citep{2000MNRAS.314...65L} or fast-mode waves \citep{2004ApJ...614..757Y}. Each of these has a different driving mechanism and different resonance kernels when it comes to CR scattering. 

In the early calculations of DSA, CRs were believed to stream along the large-scale magnetic field and were scattered back towards the shock by resonant interactions with Alfven-waves \citep[e.g.][]{Skilling.1975a}. With the realization that the magnetic field in SNRs needs to be amplified beyond the value of the large-scale field, the resonant growth of Alfven waves \citep{Skilling.1975a, 1978MNRAS.182..147B} was deemed not efficient enough \citep{1983A&A...125..249L}. An apparent solution is the non-resonant streaming instability \citep{2000MNRAS.314...65L, 2004MNRAS.353..550B}. \isBlasi{However, while in this case the turbulent field can grow to values of $\delta B\approx10 B_0$, the instability initially grows on very small scales and  the nature of the wave-particle interaction changes.}
\isBlasi{The central assumption to many studies of CR acceleration is that particles diffuse at the Bohm limit as inferred from quasilinear theory for the $\propto E^{-2}$ particle spectrum \citep[e.g.][and associated works]{2006ApJ...652.1246V, 2009MNRAS.396.2065C, 2022ApJ...936...26K, 2023ApJ...958....3D}. However, the relation between a certain energy-spectrum (and type) of magnetic turbulence and CR-scattering is neither straight forward nor well understood at the moment. Despite the existence of some evidence of Bohm-like diffusion in the presence of non-resonant magnetic field amplification as inferred from numeric simulations, this evidence is far from being conclusive.

High-resolution MHD simulations of the non-resonant instability performed by \citet{2008MNRAS.386..509R} indicate that diffusion in amplified field approaches the Bohm limit\footnote{Bohm diffusion coefficient calculated for the root mean square value of the amplified magnetic field} for particles at lower energies than those driving the turbulence growth. Further studies utilizing three-dimensional hybrid MHD–kinetic simulations implied a diffusion coefficient of $\sim4$ times larger than Bohm for particles driving the field growth, but simulations end before the saturation is reached \citep{2013MNRAS.430.2873R}. Kinetic-protons-fluid-electrons hybrid simulations imply that after a sufficient amount of time diffusion reaches the Bohm limit in the total amplified field at all energies except for the highest \citep{2014ApJ...794...47C}, but it should be understood that these simulations operate on much smaller scales and it is  unclear how to apply these results to length-scales of order the remnant size.

Finally, it should be noted that X-ray observations of young SNRs suggest that diffusion of electrons falls short of the Bohm limit in most cases \citep{2021ApJ...907..117T, 2024ApJ...973..105S}.}
While approximations are needed to tackle the DSA-problem \isrev{at the scales of real astrophysical sources}, it has to be clear that the assumption of Bohm-like scaling of diffusion and a Bohm-like turbulence spectrum constitutes the most optimistic case possible.

\subsubsection{Steady-state assumptions}
A second, common simplification is the assumption that one or more of the involved processes are operating fast enough to reach a steady state. Commonly, this is assumed to be the case for the turbulent magnetic field by assuming that the saturation-level of the non-resonant instability can be reached \citep[e.g.][]{2014ApJ...789..137B, 2018MNRAS.479.4470M, 2023ApJ...958....3D}.

Lately, a few fully time-dependent simulations \citep{2016A&A...593A..20B, 2022MNRAS.516..492B, 2021ApJ...922....7I} showed that the shock-capture time is sometimes not sufficient to allow enough growth-times for the non-resonant instability to saturate, which is in accordance with earlier analytical estimates by \cite{2008ApJ...684.1174N}. This can easily be understood as a ``chicken-and-egg-problem''. Simulations of the non-resonant instability usually assume a beam of mono-energetic high-energy particles to trigger the growth of the instability. \isrev{For example recent hybrid simulations considering a current of ions with a monochromatic initial momentum show that the saturation of the Bell instability takes place after $\sim10$ growth times when the length  scale of perturbations becomes comparable to the ion gyroradius \citep{2024ApJ...967...71Z}.} 
In reality, the particles of energy $E$ have to provide turbulence scattering particles with $E+dE$. However, scattering for particles of $E+dE$ is less efficient unless enough turbulence was build up at the right scales. This inter dependence of turbulence and maximum particle energy slows-down the acceleration process and can limit $E_\text{max}$ \citep{2016A&A...593A..20B}.

The fact that the non-resonant instability might not actually reach saturation has implications again for the \textit{scattering}-problem. The instability saturates when the scale of the driven turbulence becomes comparable to the gyro-radius of the most-energetic particles in the system. It has not been studied so far, whether the non-saturated non-resonant instability is as efficient in scattering particles as the saturated case studied by \cite{2008MNRAS.386..509R}. Steady-state approaches assume the Bohm-diffusion regime applies over the entire growth-time of the instability. 

\subsubsection{Turbulence-driving CR-flux}
Another difference in the literature, especially concerning the maximum energy, is which part of the CRs is driving the non-linear instability. The two possibilities at the extreme ends are that all CRs contribute, which requires that the field-amplification caused by escaping CRs rebuilds a quasi-planar shock geometry far upstream of the shock. This assumption is used in the approaches of e.g. \citep{2001MNRAS.321..433B, 2014ApJ...789..137B, 2020MNRAS.494.2760C} while the field-growth of \citetalias{2022MNRAS.516..492B} and \cite{2021ApJ...922....7I} is driven by particles in the precursor only, and hence lower than in the other models. \citet{2023ApJ...958....3D} calculated the maximum energy for both assumptions and finds a difference of roughly one-order of magnitude in $E_\text{max}$. 

\subsubsection{Energy fraction of injected CRs} 
Besides the particularities of turbulence growth, the actual energy carried by CRs plays the single most  important role for determining the achievable maximum energy.
Many analytic or semi-analytic approaches for the DSA problem rely on assuming the actual fraction of the explosion energy that converted to CRs. The usual assumption is that 10\% of the explosion energy get converted to CRs. However, some works assume that this fraction of energy is already converted very early on \citep[e.g.][]{2011PhRvD..84d3003M}. However, conceptually assuming a certain energy-fraction a-priori is unphysical but needed to make the problem mathematically tractable \citep{2002APh....16..429B, 
2009MNRAS.395..895C, 2011PhRvD..84d3003M, 2022ApJ...936...26K}. Recent Fermi-LAT observations of SN 2023ixf indicate, that the energy-fraction of CRs right after explosion $\leq1\%$ \citep{2024A&A...686A.254M}. \rb{Hybrid simulations on the other hand suggest injection efficiencies of the order of 10-15\% of the energy flux passing through the shock \citep{2020ApJ...905....2C}. Our simulations result in a comparable but slightly lower injection fraction, reaching of the order of 5-10\% of the explosion energy in CRs after 50kyrs \citep{2020A&A...634A..59B}.}

The actual fraction of thermal particles that participate in the DSA process is not clear and subject to ongoing studies. Our injection-fraction is tailored to observations of historical SNRs, assuming that the injection fraction is roughly constant over time and shock conditions \citep{2020A&A...634A..59B}.  

\rb{The injection efficiency that we apply here automatically ensures that the CR pressure at the shock stays below 10\% of the shock's ram pressure. Thus, the flow structure ahead of the shock is not modified by the CRs themselves and our simulation operated within the so-called test-particle limit. While we need to make this simplification due to computational constraints, we note that modifications to DSA are discussed in the community at the moment \cite[and references therein]{2020ApJ...905....2C}. They usually result in a softening of the spectral index at low energies and hence in a lower CR flux at the highest energies compared to our model. In this respect, any modification to our Ansatz softening the spectrum, will likely also reduce the maximum energy that is achieved.}

\subsubsection{Our Ansatz} In this work, we assess the magnetic turbulence by solving the transport equation for the turbulence spectrum in parallel to the transport equation of CRs \citep{2016A&A...593A..20B}. While this method was initially developed to describe Alfvenic turbulence, we include parts that are related to the non-resonant modes without adding an additional turbulence component. In any case, the diffusion coefficient varies strongly in space and time and is coupled to the spectral energy-density per unit logarithmic bandwidth, $E_w(r,k,t)$. The evolution of $E_w$ is described by
\begin{align}
 \frac{\partial E_w}{\partial t} +   \nabla \is{\cdot} (\mathbf{u} E_w) + k\frac{\partial}{\partial k}\left( {k^2} D_k \frac{\partial}{\partial k} \frac{E_w}{k^3}\right) = \nonumber\\
=2(\Gamma_g-\Gamma_d)E_w + Q \text{ . }
\label{eq:Turb_1}
\end{align}
Here, $\mathbf{u}$ denotes the advection velocity, $k$ the wavenumber, $D_k$ the diffusion coefficient in wavenumber space, and $\Gamma_g$ and $\Gamma_d$ the growth and damping terms, respectively \citep{2016A&A...593A..20B}.

The diffusion coefficient of CRs is coupled to $E_w$ by 
\begin{align}
    D_{r} &= \frac{4 v}{3 \pi }r_g \frac{U_m}{{E}_w} \text{ , }
\end{align}
where $U_m$ denotes the energy density of the large-scale magnetic field, $v$ is the particle velocity, and $r_g$ the gyro-radius of the particle.

As initial condition, we used a turbulence spectrum derived from the diffusion coefficient, as suggested by Galactic propagation modeling \citep{2011ApJ...729..106T}, reduced by a factor of ten on account of numerical constraints:  
\begin{align}
    D_0 &= 10^{28}\left(\frac{pc}{10\,\text{GeV}}\right)^{1/3}\left(\frac{B_0}{3\,\mu\text{G}}\right)^{-1/3} \frac{\text{cm}^2}{\text{s}}
\end{align}

We use a growth-rate based on the resonant streaming instability \citep{Skilling.1975a, 1978MNRAS.182..147B},
\begin{align}
    \Gamma_g &= A\cdot\frac{v_\text{A}p^2v}{3E_\text{w}}\left|\frac{\partial N}{\partial r}\right| \text{ , }\label{eq:growth}
\end{align}
where $v_\text{A}$ is the Alfv\'en velocity. We introduced a linear scaling factor, $A$, to artificially enhance the amplification. We used $A=10$ throughout this paper to mimic the more efficient amplification due to the nonresonant streaming instability \citep{2000MNRAS.314...65L, 2004MNRAS.353..550B}. Using $A$ as a scaling parameter is of course only a crude approximation of the process. However, as outlined before, there is at the moment no fully self-consistent prescription of the non-linear instability over the relevant timescales as described below. 

We found in earlier works that a value of 10 seems to be consistent with observations of historical SNRs \citep{2021A&A...654A.139B} and estimates of the growth rates operating at the early stages of CR-acceleration \citep{2018MNRAS.479.4470M}. Since we exceed the growth rate of the resonant streaming instability \citep{1978MNRAS.182..147B} by a factor of ten, the turbulent field is amplified to $\delta B\gg B_0$. 

Given this state of CR-acceleration theory in SNRs, we have to emphasize that our approach is not free of simplifying assumptions. We can only mimic the growth rate of the non-resonant instability in our fully time-dependent calculation, where we resolve the energy spectrum of the magnetic turbulence at any location, as we are not numerically able to track the return current and hence the onset of the saturation. Additionally, we assume resonant scattering of CRs on the turbulence that we drive
\begin{align}
    D(p)\propto r_g(p)\frac{B_0^2}{\delta B(p)^2}.
\end{align}
Our approach gives a more realistic description of $D(p)$, especially at larger distances from the shock compared to the usual assumption of Bohm-like diffusion everywhere. In general, our estimates for $E_\text{max}$ are within a factor of two compared to \cite{2021ApJ...922....7I} when similar ambient parameters are concerned. Further, we include cascading, which acts as a damping mechanism.

The spectral energy transfer by cascading balances the growth of the magnetic turbulence and hence the magnetic field level in our simulations. This process is described as a diffusion process in wavenumber space, and the diffusion coefficient is given by \citep{1990JGR....9514881Z, Schlickeiser.2002a}
\begin{align}
    D_\text{k} &= k^3v_\text{A}\sqrt{\frac{E_\text{w}}{2B_0^2}} \text{ . }
\end{align}
This phenomenological treatment will result in a Kolmogorov-like spectrum, if cascading is dominant. The cascading rate will depend on the level of turbulence in two regimes, since $v_\text{A}\propto B_\text{tot}$
\begin{align}
    D_k \propto 
    \begin{cases}
        \sqrt{E_\text{w}} &\text{for $E_\text{w} \ll B_0^2/8\pi$}\\
        E_\text{w} &\text{for $E_\text{w} \gg B_0^2/8\pi$} \text{ . }\\ 
    \end{cases}\label{eq:Cascading}
\end{align}
When the turbulent field starts to dominate over the background field, the cascading rate depends more sensitively on the energy density of magnetic turbulence. 

Again, this mechanism is efficiently shaping the diffusion coefficient at larger distances from the shock by redistributing scattering turbulence from scales resonant with high-energy to low-energy particles. We do not include any other damping mechanism in this work as we assume that the dense shells get photoionized by the SN flash. However, in comparison to other models, we mark already a less optimistic scenario and additional damping can only bring the maximum energy further down.

\section{Results and Discussion}\label{sec:results}
Using the wind, circumstellar shell and SN parameters in Table~\ref{tab:ProgenitorModels}, we simulated the evolution of the remnants for 20 years for each of the 6 cases. 
We focus in this paper on the evolution of the particle distribution and maximum energy in sections \ref{sec:Emax} and \ref{sec:ECR} and the amplified magnetic field in section \ref{sec:Bu}. A follow-up publication will address the emission signatures from Radio, optical, (thermal) X-rays to gamma-rays.

\subsection{Maximum particle energies}\label{sec:Emax}
\begin{figure}[htb!]
    \centering
    \includegraphics[width=0.49\textwidth]{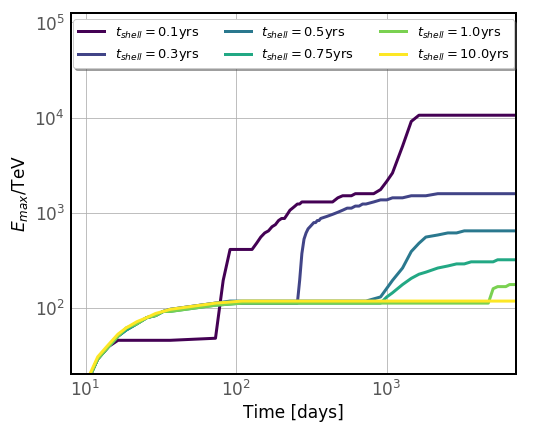}
    \includegraphics[width=0.49\textwidth]{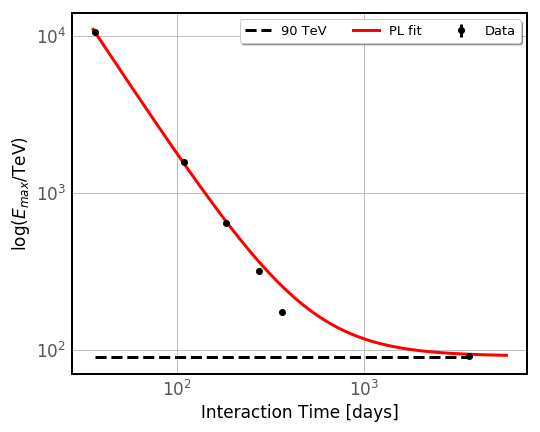}    
    \caption{\textbf{Top:} The maximum energy of protons for the simulated LBV progenitors plotted over the simulation time.
    \textbf{Bottom:} The maximum energy obtained in our simulation as a function of the onset time of shell interaction  for the LBV case (points) and the fit using the function defined in equation (\ref{eq:Emax_t}).}
    \label{fig:MaxE}
\end{figure}

To obtain the maximum energy we fitted the simulated proton spectrum below the cutoff to a power law with a fixed spectral index of $s=-2$ to obtain the spectral normalization $N_0$\footnote{Assuming that $N(E)=N_0E^{-2}$. \rb{However, this is not necessarily reflecting the true shape of the spectrum across the all energies.}}.
We then define $E_\text{max}$ as the lowest energy where
\begin{align}
    N(E_\text{max})E_\text{max}^2/N_0=1/e \text{.}\label{eq:cutoff}
\end{align}
The time evolution of the maximum Energy for our LBV simulations is shown in Figure \ref{fig:MaxE}. The shock-shell interaction decreases $E_\text{max}$ initially. For readability of the plots, we plotted $E_\text{max}(t)=\max([E_\text{max}(0),E_\text{max}(t)])$. During the interaction, substructure arises in the spectra due to injection of many particles at low energies. 
The cutoff shape changes over time and is initially super exponential with a shape following $\exp\left(-\left(\frac{E}{E_\text{max}}\right)^a\right)$, with $a>20$. The definition of $E_\text{max}$ according to equation (\ref{eq:cutoff}) avoids the influence of $a$ on $E_\text{max}$ obtained by fitting. \rb{However, we note that this treatment of the spectrum is an assumption we make to extract the maximum energy in a consistent way across all simulations. The actual particle and subsequently the emission spectra, can feature substructures and spectral breaks and strongly deviate from $s=-2$ spectra at energies below the cutoff. Observations indeed suggest very soft radio-spectra for Type-Ib and Type-Ic supernovae, indicating electron-spectra as soft as $s=-3$ \citep{2006ApJ...651..381C}. However, Type-IIP explosions which are more closely related to the cases that we focus on based on the properties of the progenitors, have been modeled with spectra closer to $s=-2.2$ and it was pointed out that cooling can significantly soften the spectra in these cases \citep{2006ApJ...641.1029C}. We discuss the emission spectra in detail in a follow-up publication.} 

The interaction of the SN with the dense shell introduces additional structure in evolution of the maximum energy for the LBV progenitors, if the  interaction takes place in the first two years after the explosion. In all these cases, the maximum energy is enhanced after the shock passes through the shell, although the jump in the maximum energy depends on the time of the interaction. In general the earlier the interaction, the larger $E_\text{max}$ becomes.
For an early interaction at $\approx0.1\,$yrs, $E_\text{max}$ surpasses $10\,$PeV. Fitting the maximum energy dependent on the interaction time (see Figure~\ref{fig:MaxE}) by

\begin{align}
 E_\text{max}=E_0\left(\frac{t}{36\,\text{days}}\right)^{-\beta}+E_1 \text{,}\label{eq:Emax_t}
\end{align}
yields $\beta=-1.80\pm0.03$, $E_0=10.8\pm1.1\,\text{PeV}$ and $E_1=90\,$TeV. 
Hence, we predict that early interactions before $\sim140\,$days lead to a maximum energy above $1\,$PeV.

\rb{This is not affected by the fact, that the case of $t_\text{shell}=0.1$yr shows more substructure in the upper panel of figure \ref{fig:MaxE} between 100-1000d. There, our ansatz using equation (\ref{eq:cutoff}) tracks a spectral feature below $E_\text{max}$ in the CR-spectrum for an intermediate time-period, which fulfills the same condition. After $1000$d, our approach finds the right cutoff energy.}

Simulations presented in \citetalias{2022MNRAS.516..492B} suggest that the non-linear instability only undergoes 3-5 growth-cycles and thus is not able to reach its saturation level, a finding supported by the work of \cite{2021ApJ...922....7I}. \rb{The reason here is that particles reside in the upstream in a region called the precursor up to a distance of $L=D(E)/v_\text{shock}$ from the shock. The time available to grow turbulence is limited to $\tau=L/v_\text{shock}=D(E)/v_\text{shock}^2$.}
We found that $E_\text{max}$ was thus limited to sub-PeV energies, whereas many other works found higher energies. In the case of interaction with dense circumstellar shells, $E_\text{max}$ is now boosted by three mechanisms:
\begin{enumerate}
    \item The shock-shell interaction slows down the shock considerably and suddenly enhances the precursor-scale $D(E)/v_\text{sh}$. The time available to grow turbulence in the precursor is enhanced. After the shock passes through the shell, the precursor scale decreases again and the shock runs through a medium with a now pre-amplified field, boosting $E_\text{max}$. 
    \item The shell-interaction boosts the CR-current and hence magnetic field amplification during and after the shell-interaction. The densities are higher for earlier interactions and hence the  $E_\text{max}$ achieved is also higher.
    \item The collision of the forward-shock with a dense shell creates reflected shocks, that can be re-reflected at the contact discontinuity and can catch up with the forward shock from behind. \rb{Such interactions with the subsequent sharp increases in the shock velocity are shown in the lower panel of Figure \ref{fig:Shells}.} These interactions enhance the forward shock's speed and slightly boosts $E_\text{max}$, as seen around a few $100\,$days for the scenario with the earliest shell-interaction. Similar effects have been described earlier for reflected shocks produced due to the the interaction of the SNR shock with the termination shocks of the progenitor star's wind-bubble \citep{2022ApJ...926..140S}. 
\end{enumerate}

In our case the shell interaction enhances the available number of growth cycles for a brief period of time while the shock accelerates after passing through the shell. This brings our estimates of $E_\text{max}$ closer to the values obtained by approaches that assume a steady state or saturation of the non-linear instability.


\subsection{Energy in comic rays}\label{sec:ECR}

We make no a-priori assumption on the fraction of the explosion energy that gets converted into CRs and fix only the fraction of the thermal plasma's particles that get injected as CRs at the shock.  The total energy in CRs is an outcome of our simulation and depends on the energy flux through the shock or, alternatively, how the shock converts kinetic to thermal energy. One can easily find that the energy passing through the shock is given by\footnote{Equation (\ref{eq:Eacc}) is strictly only valid as long as $v_\text{shock}\gg v_\text{wind}$, which is fulfilled at any time in our case.}
\begin{align}
   E_\text{acc} = \int \frac{1}{2}\rho_u v_\text{shock}^2 4\pi R_\text{shock}^2 v_\text{shock} \text{d}t \propto t^{3a-2} \label{eq:Eacc},
\end{align}
where $\rho_\mathrm{u}$ is the upstream density, $v_\text{shock}$ the shock velocity, $R_\text{shock}$ the shock radius and $a$ the expansion parameter, where $R_\text{shock}\propto t^a$.
The point where $E_\text{acc}$ roughly equals the explosion energy marks the transition to the Sedov stage of the remnant's evolution. In the first few years, $E_\text{acc}$ is only small fraction of the $10^{51}\,$erg of explosion energy.

\begin{figure}[htb!]
    \centering
    \includegraphics[width=0.49\textwidth]{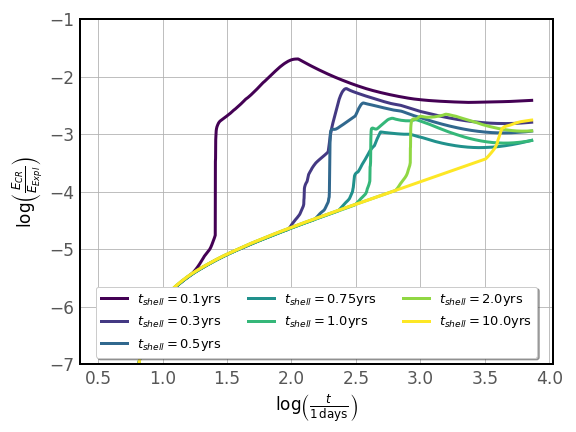}
    \caption{The total energy in CRs in units of $10^{51}\,$erg for the LBV-interaction simulations with different shell radii, as a function of time.}
     \label{fig:Ecr}
\end{figure}

Figure \ref{fig:Ecr} shows the evolution of the total energy in CRs. Within the first day, the energy in CRs rises fast and then continues to increase roughly as $\propto t$, following from equation (\ref{eq:Eacc}) as our initial expansion parameters is $a\sim1$. However, the interaction with the shells breaks this power-law dependence, greatly increasing $E_\text{acc}$ and hence the energy in CRs.

The early-interaction case reaches a peak energy-fraction of $\sim2\,$\%, marking the case with the highest conversion efficiency among our Type-IIn cases. The increase of available energy and the increased total energy in CRs supports the strong increase of $E_\text{max}$. \rb{We note though, that the yield in reality might be lower than our predictions and our predictions have to be considered as upper-limits. The medium around the SN can feature strong inhomogeneities itself or the explosions can be asymmetric as observed for many core-collapse explosions. This reduces and/or smears the effects resulting from the shell-interactions that we discussed in this paper over longer time-periods.}

\subsection{Magnetic field evolution}\label{sec:Bu}
\begin{figure}[htb!]
    \centering
    \includegraphics[width=0.49\textwidth]{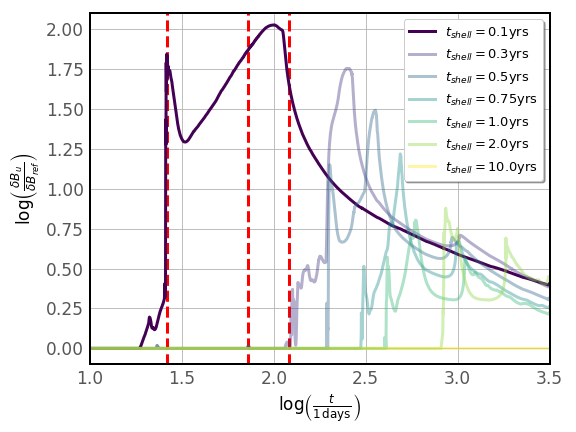}
    \caption{The solid lines show the log-ratio of the amplified magnetic field in the immediate upstream relative to the field of a simulation without any shell interactions. The vertical dashed red lines mark from left to right the times of the first shock-shell interaction, the time of when the shock reached the peak-density of the shell and the time when the shock is propagating in the free wind again for the simulation with the earliest shell interaction (strong line). The pale lines show the behavior for later shell interactins.}
     \label{fig:Bu}
\end{figure}

Crucial for the acceleration of particles is the evolution of the turbulent field. We calculated the component of the turbulent field from an integral over the turbulence spectrum and plotted the time-evolution in Figure \ref{fig:Bu}.

Figure \ref{fig:Bu} shows the amplified field relative to the amplified field of a simulation without a shell-interaction. 
As shell interaction begins (the first red line from the left marks the point where the pre-shock density increased by $3\times$), there is an initial surge in the field immediately downstream of the shock.
This is partially due to the increasing CR current, and partially arises because particles of the highest energies are no longer confined and can escape the remnant from deep downstream.
This escape creates a limited spike and the field drops again as the importance of the particles escaping from deep-down fades compared to particles that get injected freshly into the upstream.

After this first spike, the field increases again till after the peak of the downstream density (marked by the middle red-line in Figure \ref{fig:Bu}) is reached. The period of rising field is caused by the steady increase of the CR-current due to the increasing density and the fact that the surge of particles escaping during the onset of the shell-interaction pre-amplified the upstream field.

The upstream field strength starts to fall shortly before the shock completely passed the dense shell, when the shock starts to accelerate again. At its peak, the field is roughly a factor of 100 higher than compared to the case without a shell interaction. 
Many CRs that were injected during the shell interaction are still present and can drive field growth, but the field decays over time along with the density of injected particles.
At the time when $E_\text{max}$ is reached after roughly 100 days, the upstream field is still a factor of 5 higher compared to the case without a shell interaction.
The same structure of field growth can be seen in the scenarios with a later shell-interaction, although the upstream field-strength at the end of the simulation and during the peak is lower in all other scenarios.

\subsubsection{Comparison to other models}

As described in section \ref{sec:Turb}, there is no model available at the moment that captures all the underlying physics of CR acceleration in a satisfactory manner. As a consequence, our predictions reflect the assumptions and shortcomings of our ansatz.

In general, we obtain maximum energies that are at the low end of the models in the literature, with values of $\approx 100$~TeV before the shell-wind interactions. 
This is considerably lower than the values obtained\footnote{The values given here are for a mass-loss rate of $10^{-2}\mathrm{M}_\odot \,\mathrm{yr}^{-1}$. If the work used a different mass-loss rate, we adopted a scaling of $E_\text{max}\propto\sqrt{\dot{M}}$.} by \citet[][21~PeV]{2013MNRAS.431..415B}, \citet[][8~PeV]{2018SSRv..214...41B}, \citet[][60~PeV]{2018MNRAS.479.4470M}, \citet[][7.5~PeV]{2021ApJ...922....7I} and \citet[][2.4-60~PeV]{2023ApJ...958....3D} due to the fact that we use an injection fraction of CRs that is on the low end of the range assumed in the above-cited works. Before the shell-interaction only 0.01\% of the explosion energy is converted into CRs. Additionally, the fact that we are not considering the saturation of the non-resonant instability and that the number of turbulence growth-cycles is limited, further limits $E_\text{max}$. Our model is most comparable with the work of \cite{2021ApJ...922....7I} but, on account of turbulence cascading which acts as a damping mechanism, we achieve an $E_\text{max}$ that is about a factor of $\approx2$ below the values obtained by \cite{2021ApJ...922....7I}. Both works also only consider the driving of turbulence by CRs in the precursor.

The early shock-shell interactions change a few of the points above. Most importantly, the deceleration of the shock, that enhances CR-escape followed by an acceleration that lets the shock run through a medium where escaping CRs drove turbulence,  greatly enhances $E_\text{max}$ by basically allowing a larger number of growth-cycles. As a consequence, our maximum energies become comparable to more optimistic models like \cite{2013MNRAS.431..415B} and \cite{2018SSRv..214...41B}. Despite the fact that we are not switching off the driving of turbulence when the (analytical) saturation limit is reached, \is{efficient} cascading 
continues to limit $E_\text{max}$ as the damping rate of turbulence starts to scale with $E_w$ when $E_w \gg B_0^2/8\pi$ (see also equation \ref{eq:Cascading}). Further, the interactions boost the number of CRs and enhance the energy converted to CRs greatly, reaching up to $\sim2$\% after the interaction. This of course provides more energy that can eventually be converted to turbulence at larger scales, essential for the enhancement of $E_\text{max}$.

\section{Conclusions}\label{sec:conclusions}
We performed numerical simulations of particle acceleration in very young SNRs expanding in dense circumstellar media featuring dense shells created by LBV progenitors, solving time-dependent transport equations of CRs and magnetic turbulence in the test-particle limit alongside the standard gas-dynamical equations for CC-SNRs. We derived the CR diffusion coefficient from the spectrum of magnetic turbulence that evolves through driving by the CR-pressure gradient, as well as cascading and wave damping.

We found, that the maximum proton energy that we observe in our simulations exceeds PeV energies when the shock interaction with a shell of about $2M_\odot$ takes place prior to $\approx140\,$days. Later interactions still boost $E_\text{max}$ but not beyond the PeV frontier. \rb{The maximum energy as function of the interaction-time with the shell follows a power law with of the form $E_\text{max}\propto t^\beta$ with $\beta=-1.80\pm0.03$.} The increase of the maximum energy is driven by a combination of  (i) the increased CR current due to the increasing ambient density during the interaction, and (ii) the increase and then decrease of the precursor scale caused by the temporary drop in the shock velocity during the shell interaction. \isrev{It should be emphasized that such a scenario which reaches PeV energies is limited to very young SNRs and is plausible only for a small subset of SNRs. In this sense our results agree with previous works arguing that SNRs are unlikely to be responsible for most of CRs around the \emph{knee}.}

The upstream magnetic field is largely boosted when the shock passes through dense material and at the same time the fraction of the explosion energy that is processed through the shock increases. Hence, a much larger fraction of the explosion energy is available for particle acceleration and consequently will also boost the gamma-ray emission, which we address in a follow-up publication.

\section*{Data availability}
The data underlying this article will be shared on reasonable request to the corresponding author.
\section*{Acknowledgements}
RB and JM acknowledge funding from an Irish Research Council Starting Laureate Award. 
JM acknowledges funding from a Royal Society-Science Foundation Ireland University Research Fellowship.
This publication results from research conducted with the financial support of Taighde \'Eireann - Research Ireland under Grant numbers 20/RS-URF-R/3712, 22/RS-EA/3810, IRCLA\textbackslash 2017\textbackslash 83.
We acknowledge the SFI/HEA Irish Centre for High-End Computing (ICHEC) for the provision of computational facilities and support (project dsast026c). IS acknowledges funding from Comunidad de Madrid through the Atracción de Talento “César Nombela” grant with reference number 2023-T1/TEC-29126.
\rb{And finally we thank Martin Pohl and Pasquale Blasi for the useful discussions and suggestions.}

\bibliographystyle{aa}
\bibliography{References}

\end{document}